# Research on Early Warning and NB-IoT Real-time Monitoring System for Radiation Source Shedding of Gamma Flaw Detection Machine

ZHANG Zheng-yang[1], LIU Zhi-hui[2]，ZHANG Rui[3]，HE Rong-hua[4*], WANG Zhe[2]

(1.Beihang University, Beijing 100083, China. 2. Beijing Radiation Safety Technology Center, Beijing 100089, China. 3.University of British Columbia，Vancouver,V6T 1Z41. 4.China Institute of Atomic Energy, Beijing 102413, China. *corresponding author)

**Abstract:** The system takes the embedded system single chip as the core, and organizes the gamma ray induction module, keying switch, radiation source braid locking mechanism, on-site alarm equipment, NB-IoT communication module, GPS positioning system and other related equipment to realize real-time operators warning and remote alarming to the monitoring platform. Thus, timely management of the fallen radiation source can be realized. What's more, position of radiation source braid is monitored by radiation source braid locking mechanism, gamma-ray induction module and keying switch. This idea solves the bottleneck problem of difficulty in transmitting real-time remote monitoring signal by NB-IoT technology. Single-chip microcontroller is innovatively embedded around the radiation source to monitor whether the source falls off. As a result, when part or all of the source braid are not returned to the storage location of the flaw detector, the gamma ray induction module, keying switch, and radiation source braid locking mechanism remind the operator of the dangerous situation. At the same time, NB-IoT transmission system alarms companies and regulators of safety risks. In conclusion, radiation source can be found timely when it falls off, to avoid radiation accident. Two bottlenecks related to installing GPS positioning system can be solved, that the GPS positioning system is installed on the flaw detector, and the radioactive source is still unable to be controlled, and that the GPS positioning system requires power to transmit the signal back to the regulatory platform or to the platform of the flaw detection enterprise. This article's ideas can solve this problem through the advantages of NB-IoT.

## Introduction

γ-ray flaw detection is widely used as non-destructive testing technology. It is mainly used to detect the integrity of equipment and structure, such as containers, pipes, welds, castings, etc. It is also widely used in metallurgy, petrochemical, manufacturing, installation and other industries. Gamma-ray flaw detectors have the following features: 1. Low price, large detection thickness and strong penetrating ability; 2. Small volume, no need to provide water and electricity supply, and therefore, it is especially suitable for field operations; 3. High efficiency, such as being able to carry out circumferential exposure and panoramic exposure to the detection container; 4. No wearing parts, low equipment failure rate; 5. Being able to be continuously operated without being affected by external conditions such as temperature, pressure, magnetic field, etc. (The shooting conditions can be determined through simple calculations, and the shooting process is stable.); 6. Good operability (The gamma flaw detection container can be used to store gamma rays. The sealed source is in a safe position in the container when the source is not working. When exposing with the source, the machine sends the source from the safe position to the irradiation head through the source conduit.). It has the irreplaceable characteristics of other flaw detection devices. [1] Therefore, many developed countries widely use γ-ray flaw detectors. With the rapid development of modernization and the improvement of product quality requirements in China, the usage of γ-ray flaw detectors is increasing, obtaining good technical results and economic benefits.

However, the use of γ-ray flaw detectors still has huge radiation hazards: due to the lack of safety management systems, equipment quality problems and the technical problems of users, China's γ-ray flaw detectors have frequent accidents. There have been many accidents in Guangzhou and other places in which radioactive sources were lost, stolen, and people were exposed [2]-[5].

At present, the radioactive sources used by γ-ray flaw detectors are mainly iridium-192, selenium-75, cesium-137, cobalt-60, etc., which are Class II radioactive sources and high-risk sources. If detectors are improper used or poor managed, it will cause serious radiation accidents. According to the relevant legal provisions, out-of-control phenomena of lost or stolen Class II radioactive sources are treated as major radiation accidents. Because the working time is in the midnight, the workplace is in the wild, and the gamma flaw detection has great potential safety hazards during transportation and storage, the γ flaw detection source has high fluidity and high risk. The main radiation risk faced by nuclear technology utilization is the fall-off of the gamma defect detection source. In the accidents, it was not found that the radiation source was lost to the publics, causing radiation damage to human health, radiation accidents, and social panic.

Currently, regulatory authorities cannot understand the dynamic usage of gamma sources, and supervision is in a passive state. It is envisaged that installing a GPS positioning system to real-time track the exact position of the gamma flaw detector and grasp the dynamics of the gamma flaw detection source in the area can reduce accidents. However, in the field operation of γ flaw detection, most of the working conditions are in remote areas has no power supply. The data link signal is weak, and online monitoring cannot be achieved. The main issue is that the GPS installed on the flaw detector, which is actually monitoring flaw detector, and it grasp whether the gamma flaw detector is in place. Even if a GPS positioning device is added to the gamma flaw detector, the purpose of dynamically monitoring the gamma source cannot be achieved.

This paper introduces the narrowband Internet of Things technology. The narrowband Internet of Things (a cellular data network) is generally built by communication operators while building other types of cellular data networks. Users don't need to set up their own network but to purchase narrowband IoT services. Different from the commonly used 3G or 4G cellular communication methods, the narrowband Internet of Things uses a narrower frequency band, sacrifices transmission speed, in exchange for extremely low power consumption, and reduces the minimum requirements for the signal-to-noise ratio of the communication system.

These characteristics of the narrowband Internet of Things are just suitable for applications in the field of flaw detection. As the location of the flaw detector will change with the construction work process, it is unrealistic to require the flaw detection unit to form a network in advance at the construction site before using the flaw detection machine. However, the narrowband Internet of Things is networked by communication carriers, and the construction unit or manufacturer of the flaw detector only needs to purchase the services of the communication carriers. Thus, communication can be realized in the places covered by the network signals of all operators. In addition, the use of flaw detectors is often in the wild or in new buildings. Although there will be signal coverage of communication, due to the influence of distance, obstacles and multipath effects, the signal-to-noise ratio of the channel is often relatively low. But the NB-IoT technology can tolerate an additional 20dB signal-to-noise ratio, which solves the problem that the existing flaw detection machine using GPRS communication technology often encounters a communication failure. The existing scheme of flaw detectors using GPRS communication technology causes the flaw detectors to be charged regularly to ensure the normal operation of the positioning module. The low-power consumption nature of the narrowband Internet of Things makes the flaw detectors unnecessary to replace or charge the battery throughout its life cycle.

The purpose of this study is to combine the use of NB-IoT technology [6]-[9] and the use of $\gamma$-ray flaw detection online monitoring[10]. ① NB-IoT narrowband Internet of Things technology realizes remote positioning and dynamic online monitoring of gamma radiation sources in field flaw detection. ②The embedded system displays whether the gamma flaw detection source is in place and automatically alarms in real time. The ecological and environmental protection departments can implement comprehensive and effective online monitoring of gamma flaw detectors and gamma radiation sources, thereby greatly improving the supervision level and efficiency of the flaw detection industry.

1. **System architecture**

This system takes embedded system single-chip microcomputer as the core, and organizes related devices such as γ-ray sensor module, key switch, source braid lock mechanism, field alarm equipment, NB-IoT communication module, GPS positioning system, etc. Combined with the source braid locking mechanism, the system uses a mechanical method to detect the position of the source braid to realize the function of the whole system.

1.1 Hardware

The overall internal hardware design structure of the flaw detector is shown in Figure 1:

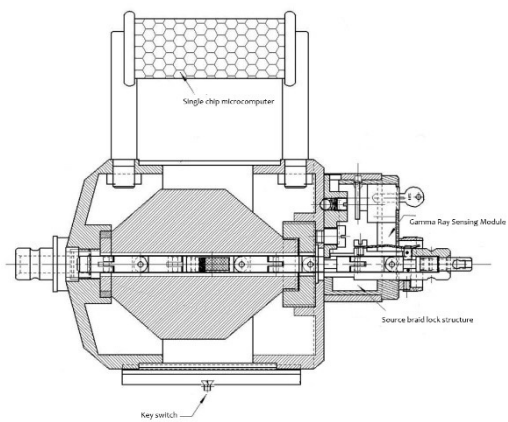
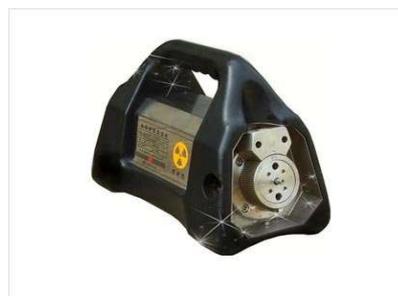

Fig1　Internal profile of the detector　　　Fig2　The appearance of the detector

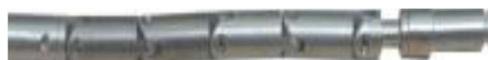
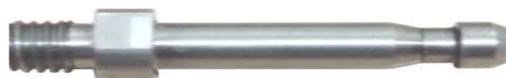

(a)　　　　　　　　　　　　　　　　(b)

Fig3　Source braids (a) and quick joints (b)

Figure 2 illustrates a straight pipe flaw detector currently in use. Figure 3 illustrates the source pigtail on the left with the flaw detection source installed. When not working, it is stored at the location inside the flaw detector. When flaw detection is required, unscrew the quick connector on the right as shown in the figure, then connect the quick connector to the source pipeline, and send the source pigtail to the part that needs flaw detection.

The existing operation mode is that the source pigtail locking mechanism only has the locking function at the recovery position. The switch ring of the safety lock mechanism can still be rotated normally to close position even if the shedding has occurred when the source braid is recovered. In fact, when the radioactive source has not returned to the safety shielding position, it shows false appearance that the radioactive source has been recovered. Therefore, misjudgment is caused, and an irradiation accident occurs. The existing structure has the technical problem of not being able to sense whether the source braid is returned.

This study addresses this dilemma. Its technical solution designs a mechanism: a door-lock-like structure that can indicate that the source braid has not been retracted. When the source braid falls off and is not retracted to the safety shielding position, the locking groove of the switch ring is inserted into it, so that the switch ring cannot be rotated to the closed position. Therefore, it reminds the inspection staff that the radioactive source braid is not normally returned to a safe position, to warn the inspection personnel to start using the radiation measuring instrument to find the inspection source, so as to avoid the accident of losing the source.

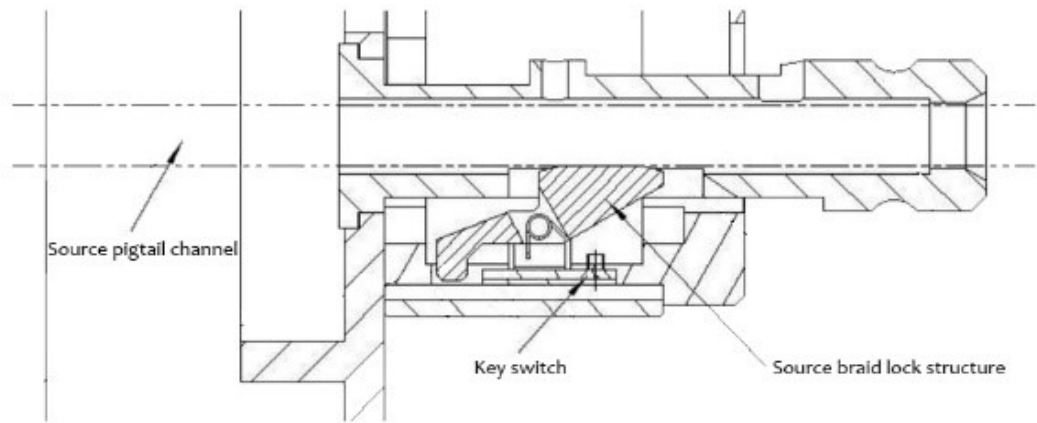

Fig4   A diagram of the source braid lock mechanism

1.1.2 Gamma ray sensing module
The technical scheme designs a gamma ray sensing module. It is used to show that the source braid of the gamma ray NDT machine is outside the flaw detector. The location of the γ-ray sensing module is shown in Figure 4.

In view of the bad working conditions of the flaw detection machine, and brutal operation of the user occurs from time to time, anodized aluminum is wrapped in the outermost layer of the structure to protect it.

The γ-ray sensing module is embedded near the retracted position of the source braid (at the handle on the right side of Figure 1), and the above device is protected with anodized aluminum + thin lead sheet in the position contact with the source braid. The barbaric operation of personnel may damage the function of the γ-ray sensing module. The thin-layer lead sheet has the effect of shielding the γ impurities in the depleted uranium tank and visible light exposure. When the value of the radiation dose received by the aforementioned gamma ray sensing module is higher than a preset threshold, it is confirmed that the source braid has protruded outside the flaw detector and entered the source duct. It is realized by photoelectric effect principle.

1.1.3 Keyed Switch
This technical solution designs a key switch for electronically sensing where the flaw detector is located, the retracted state or the operating state.

By setting two switches, the first switch is set at the radiation source retracting position or the source braid lock mechanism, and the second switch is set outside the bottom casing of the flaw detection machine. The pop-up switch requires external force to keep it in the pressed state. When the external force is removed, the switch lifts itself. The push-in and pop-up of the switch correspond to the on and off states of the circuit, respectively.

The first switch is designed to be located in the source retracting position or in the source braid locking mechanism, and is used to detect whether the source has been fully retracted to a safe position in the depleted uranium tank. When the radiation source has been completely retracted, the external force will press the first switch, and the circuit is turned on.

The first switch also has a servo circuit for enabling the embedded system to detect the state of the switch. The implemented circuit has multiple combinations, and only one is listed in this embodiment. The specific circuit is shown in Figure 5.

When the switch is turned off, the detection point detects a low voltage, and the embedded system learns that the radiation source is not in a safe position; when the switch is closed, the detection point detects a high voltage, and the embedded system learns that the radiation source has returned to a safe position. The low voltage in the logic of the circuit should correspond to an unsafe state to avoid the false report of a safe state due to a circuit failure.

The designed second switch is arranged outside the bottom casing of the flaw detector. When the flaw detector is placed on the ground, the second switch is pressed and the circuit is turned on.

The second switch also has a servo circuit for enabling the embedded system to detect the state of the switch. The implemented circuit has multiple combinations, and only one is listed in this embodiment. The specific circuit is shown in Figure 6.

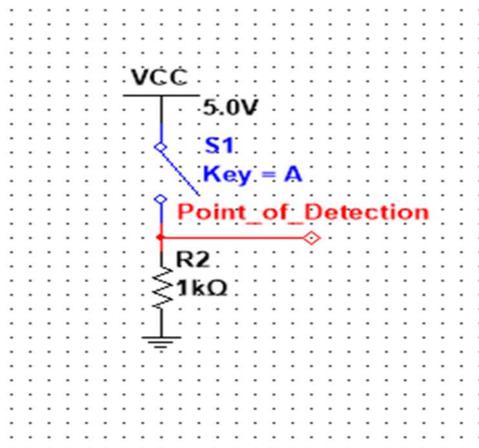
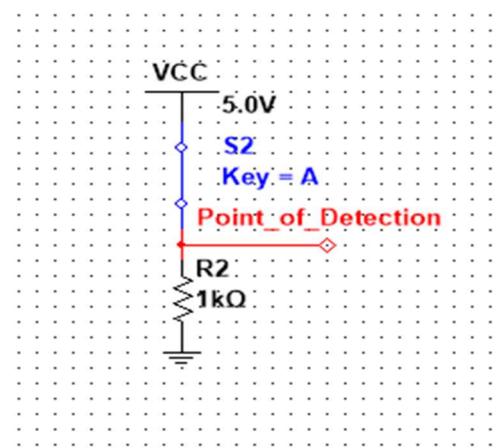

Fig5　Servo circuit for the first switch　　　Fig6　Servo circuit for the second switch

When the switch is turned off, the detection point detects a low voltage, and the embedded system learns that the flaw detector is lifted off the ground; when the switch is closed, the detection point detects a high potential, and the embedded system learns that the flaw detector is placed on the ground or other solid surface. Preferably, the low potential in the circuit logic should be caused to correspond to the lifted state, so as to avoid a false report of a safe state due to a circuit failure.

1.1.4 GPS positioning system

In this system, the GPS positioning system module is directly installed on the flaw detector or integrated directly on the circuit board of the single-chip microcomputer; the single-chip microcomputer directly reads the data packets output by the GPS positioning system module and performs data processing inside the single-chip microcomputer.

1.1.5 NB-IoT communication module

As for the hardware structure, the highly integrated NB-IoT communication module produced by the NB-IoT communication module manufacturer is connected to or integrated into the single-chip microcomputer circuit board. Software and hardware interfaces of the NB-IoT communication module designed by the manufacturer when leaving the factory communicate with the microcontroller. When designing, it is no longer necessary to consider the internal hardware structure of the NB-IoT communication module.

1.1.6 Field alarm equipment

On-site alarm equipment uses sound and light alarm equipment such as alarm bells, siren, lights, etc., to timely and prominently send an alarm signal to alert the site staff and the surrounding public. The accident can be handled at the scene.

1.2 software part

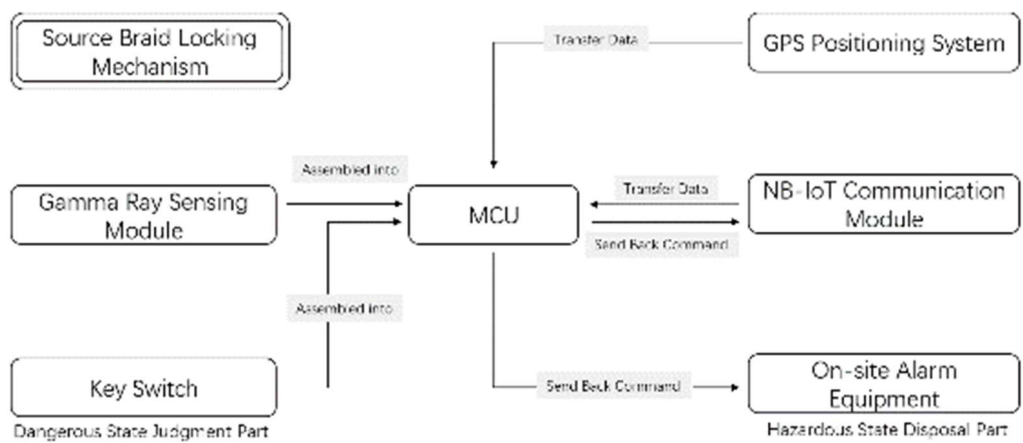

Fig7　The architecture of the hazard state judgment system and disposal system

### 1.2.1 Internal System Architecture of Flaw Detector

This system uses an embedded single-chip microcomputer system as the core of the entire software system, as the "brain" of the system, to sense the state of the flaw detector, to determine whether the system is in a dangerous state, and to responds accordingly.

In this study, the mechanical control switch of the source braid locking mechanism was used to monitor whether the source braid of the flaw detector returned to the safe storage position in the depleted uranium tank of the flaw detector. Using a gamma-ray sensing module, the reaction source braid is extended. The key switch electronically senses the retracted state of the radioactive source and the operating state of the flaw detector. Three technical methods, integrated into the embedded system's single-chip microcomputer to achieve early warning, enable field operators to find the cause of the failure in a timely manner and verify the exact location of the source braid. The γ-ray sensing module, the key switch, and the source braid lock mechanism are used as the dangerous state judgment part.

At the same time, the NB-IoT narrowband IoT technology is used to transmit the early warning signals to the company's management platform and environmental protection supervision platform. NB-IoT communication module, on-site alarm equipment, GPS positioning system and other related equipment work together to deal with dangerous conditions.

This study highlights the innovative and optimized optimization of the interaction between embedded system modules, so that the system has the characteristics of low energy consumption and long life on the premise of safety and accuracy. The following describes the working process of the software with the help of the description of the working scene of the flaw detector, and reveals the significant advantages that this design solution does not have in other solutions.

When the flaw detector is not in use, each system is in a dormant state, and the system power consumption is very low to save battery power.

### 1.2.2 γ-ray sensing module

When the radiation source protrudes from the flaw detector, a current signal is obtained at the port of the gamma-ray sensing module. Use this current signal to wake up the microcontroller. Because the γ-ray induction module is designed using the principle of photoelectric effect, when there is no γ-ray entering, the circuit is disconnected, there is only voltage and no current across the module, that is, no energy consumption. Only when the flaw detector is working, that is, when the source braid is extended and the gamma rays enter the module, a current will exist in the module. Therefore, this solution uses this current signal to wake up the microcontroller, which effectively reduces the power consumption of the system.

The principle is shown in Figure 8.

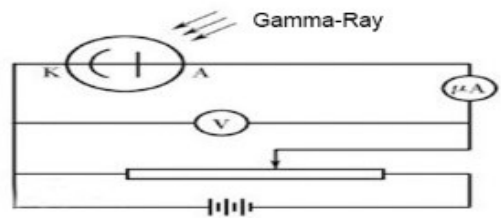

Fig8 Principles of the Gamma-Ray Sensing Module

The photoelectric effect principle is used to detect the presence of gamma rays. When γ-ray photons are irradiated on the cathode plate, the electrons outside the nucleus of the atoms in the cathode plate absorb the energy of the γ-ray photons and are excited, free from the nucleus of the atomic nucleus, and become free electrons. Electrons accelerate toward the anode under the action of an electric field formed by the power source, and are eventually absorbed by the anode. This process makes the closed cathode and anode in the original circuit form a closed circuit, and communicates with the other circuit components in a closed loop, so that the single chip microcomputer can detect the current in this circuit.

However, because depleted uranium is used as a shielding material for the radioactive source, the gamma rays emitted by the depleted uranium can also cause the above-mentioned photoelectric effect, which causes the single-chip computer to incorrectly judge the working state of the radioactive source. To solve this problem, a thin layer of lead sheet is wrapped around the gamma-ray sensing module. Since the radiation dose of the flaw detection source is very different from the radiation dose of depleted uranium, a thin-layered lead sheet with a proper thickness can block the radiation dose of the depleted uranium, but at the same time allow the radiation dose of the remaining flaw detection source to pass through and enter the cathode, causing a photoelectric effect. In this way, the single-chip microcomputer can accurately identify the working status of the radioactive source.

1.2.3  the source braid locking mechanism
When the radiation source falls off, the source braid locking mechanism is locked, and the quick connector and the connected source conduit are prohibited from being removed. The key switch located at the position of the source braid lock mechanism recognizes the locked state of the source braid lock mechanism, and feeds back this state to the microcontroller At this time, the single chip microcomputer learns that the flaw detector is in a dangerous state.

1.2.4  Keyed Switch.
If at this time, the operator forcibly picks up the flaw detector, and wants to take the equipment away from the scene, the key switch set outside the bottom casing of the flaw detector pops up, and recognizes this dangerous action of the operator. The dangerous state detection switch transmits the current state to the single-chip microcomputer through the servo circuit, and the single-chip microcomputer determines the current dangerous state according to the preset logic.

In the implementation process, the circuit logic corresponding to the pressing of the switch and the conduction is shown in Table 1.

Table 1 Relationship between radioactive source location and alarm states

| Radioactive source location | First switch truth | Flaw detector location | Second switch position | Whether to call the police |
|---|---|---|---|---|
| Take back | high | On the ground | high | No |
| Take back | high | Lift up | low | No |
| Take out | low | On the ground | high | No |
| Take out | low | Lift up | low | Yes |

1.2.5  On-site alarm equipment

At this time, the dangerous state judgment part has confirmed that the flaw detector is in a dangerous state, and the single chip microcomputer will instruct the dangerous state handling part to start alarming.

The single chip microcomputer first turns on the power of the on-site alarm device. The on-site alarm equipment issued an audible and visual alarm signal to remind the on-site staff that there is a problem now.

1.2.6  GPS positioning system

The single-chip microcomputer wakes up the GPS positioning system and obtains the positioning information of the flaw detector from the system in real time.

It is easy to purchase highly integrated single-chip microcomputer modules on the market. The single chip microcomputer directly reads the data packets output by the GPS positioning system module, and performs data processing inside the single chip microcomputer. It does not involve the hardware architecture and circuit design of the GPS positioning system module.

1.2.7  NB-IoT communication module

The single-chip microcomputer instructs the NB-IoT communication module to send an alarm message, and sends the positioning information of the flaw detection machine in real time, and informs the senior management of the flaw detection enterprise or the government supervision department about this alarm information. At the same time, it receives instructions from the flaw detection machine monitoring cloud system.

When the microcontroller is sleeping, the NB-IoT communication module can also obtain instructions from the server, forcing the microcontroller to wake up, and execute related instructions.

From the above description of the system's working process, it can be seen that in the entire real-time monitoring system, devices with large power consumption, such as GPS positioning systems and field alarm devices, are turned off when no alarm is required, and do not consume battery power.

Devices with low power consumption, such as single-chip microcomputers and NB-IoT communication modules, are in standby state when the flaw detector is not in use, with extremely low power consumption, which minimizes the power consumption of the real-time monitoring system and extends battery life; And monitoring switch, using analog signal detection technology based on voltage and current, there is no power consumption in standby state, and power consumption is very low when in use.

Through these means to reduce the overall system energy consumption, the excellent characteristics of the flaw detector without replacing the battery throughout the entire life cycle of the flaw detector are achieved.

On the other hand, the source braid lock mechanism is designed using the principle of pure mechanics. Even when there is a failure in the single-chip computer system or circuit system or the battery power is exhausted due to over-life use, it can still work normally to lock the connector, and prohibit fast connectors The way the connected source conduit is removed effectively reminds the operator of the danger.

1.2.8 Cloud Detection System for Flaw Detectors
The signal transmission system is shown in Figure 9.

This article uses NB-IoT technology to transmit: The data transmission is shown in Figure 7: ①Front-end monitoring system: Monitor and collect process and status information on the place where the radioactive source is located. ②Keyed gamma flaw detector: It is used
to detect whether the radioactive source is returned to the source tank monitoring agency every time. If it does not return to the source tank, it will send an alarm signal. ③Monitoring platform: comprehensively and accurately understand the status of the location of the radioactive source and other information. Automatic alarm for abnormal conditions.④Monitoring Center: Link to the monitoring end of the radioactive source through wireless transmission to realize whether the radioactive source is off-line and continuous status monitoring.

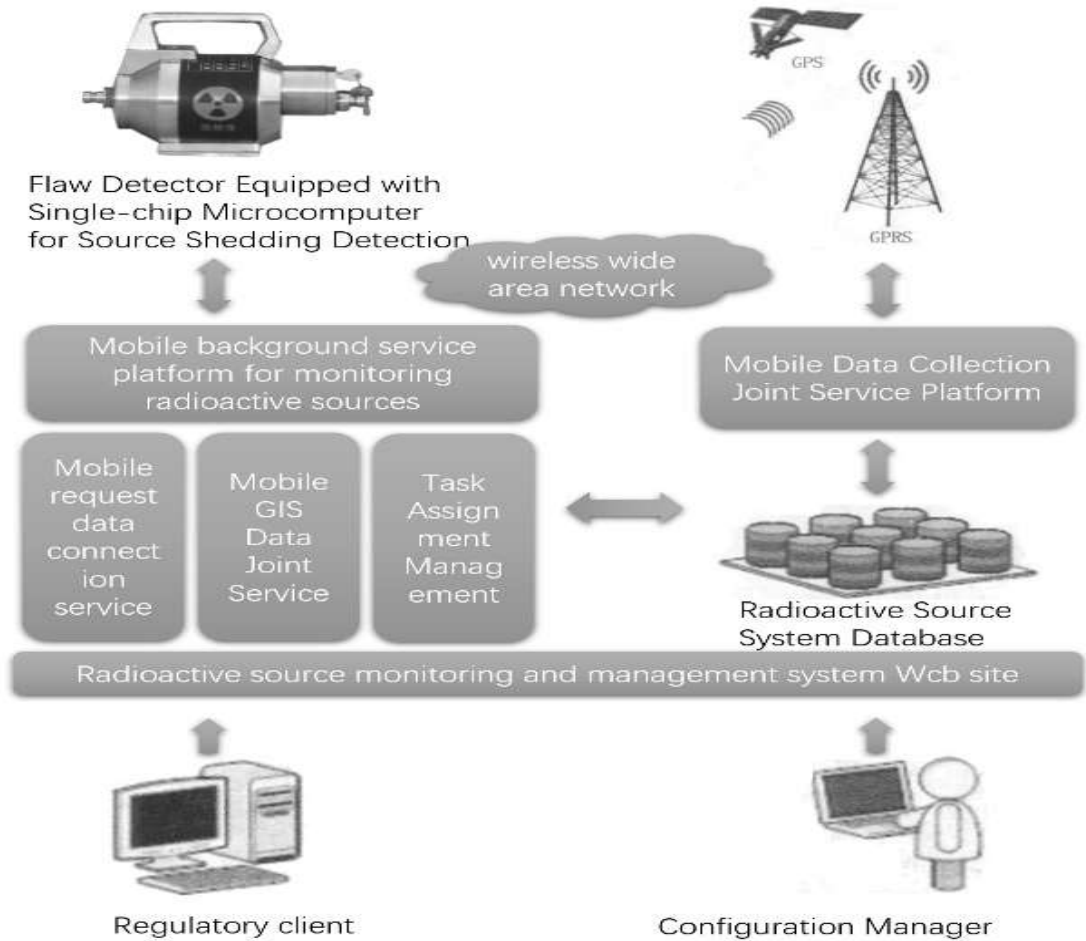

Figure 9 Structure of the transmission system

According to the result of logical judgment, if an alarm is required, the method in the previous table is used to alarm through the NB-IoT system.

**2. The advanced nature of the same type of technology comparison**

At present, the radiation supervision department requires that GPS be installed on the gamma flaw detector. The bottlenecks are: 1. The GPS positioning system only tracks the flaw detector and cannot determine whether the radiation source has fallen off. 2. GPS requires power to transmit signals back to the supervision platform and flaw detection enterprise platform, while mobile flaw detection is basically in a relatively remote place. Without power supply, the base station coverage of cellular data networks is relatively poor. This article uses three technical methods: 1. source braid lock mechanism; 2. γ-ray induction module; 3. keying (dangerous state detection) switch. The combination of the three technologies with the NB-IoT narrowband IoT transmission technology has solved the problem of tracking the source of gamma flaw detection and broke the current bottleneck.

Aiming at the problem of loss control of γ-detection radioactive sources, this paper uses the current advanced NB-IoT technology to solve the problem of transmission of status signals. Solve the bottleneck problem that the current flaw detection field cannot monitor the loss and loss of radioactive sources. After the gamma radiation source is lost, it will send an alert to the company's platform or the regulatory department's platform as soon as possible, and if necessary, wake up the GPS positioning system. It can send technicians to verify the accuracy of the warning in time. Once the radiation source falls off, the organization personnel can retrieve the radiation source. The first time., The narrowband IoT technology is introduced to the flaw detection industry for early warning.

The mechanical switch of the source braid lock mechanism designed in this article is installed in the quick connector of the flaw detector in the production stage. A door-like structure is installed. When the source braid does not completely return to the safe position of the flaw detector, the door lock mechanism cannot be closed to alert the flaw The staff were looking for a source braid that could be missing.

The γ-ray induction module designed in this paper uses the principle of photoelectric effect and is installed near the source retracting position outside the flaw detector, showing that the source braid has left the flaw detector storage location, and the quick connector cannot be closed outside the depleted uranium tank. Remind flaw detection staff to find the source of flaw detection on the spot.
In this paper, a binary logic method key switch is used for the first time to identify the radioactive source position outside the radioactive source sleeve to identify whether the radioactive source has fallen off and monitor the dangerous state.

These three technical means are all advanced technologies that have not yet been used through a new search.

These three technical methods and the application of NB-IoT technology in early detection of flaw detection sources are all proposed for the first time, and are innovative and technologically advanced.

This article addresses the multi-risk problem of γ flaw detection, and will take advantage of the four major advantages of the NB-IoT cellular narrowband IoT: The first is wide coverage (deep coverage is better), and the network has about 20dB gain under the same conditions.The second is the large connection, a single cell supports 50,000 to 100,000 links.The third is low cost [11], which greatly reduces the cost of chips. Fourth, the power consumption is low, and the standby work is up to 10 years, which is equivalent to the life of the flaw detector. NB-IoT related technologies are integrated into the monitoring system of radioactive sources. Real-time recording of the radioactive source status is achieved, and an alarm is issued when an abnormality occurs. The system's entire process records can improve the safety of the use of radioactive sources, save time and effort, and greatly protect the lives of employees and the people. The application of NB-IoT is getting wider and wider, but the idea of applying it to the flaw detection industry is proposed .

## 3. Conclusion

This article uses the NB-IoT [12]-[16], which has been applied, to integrate the technologies of the two disciplines of NB-IoT and the flaw detection industry, and introduce NB-IoT technology. Induction module (photoelectric induction theory is put into practice) and key switch detector technology. It has solved the bottleneck problem that the flaw detection industry cannot track the flaw detection source at present and cannot find the disposal in time after the loss.

Because NB-IoT has stronger network coverage, large access capacity, low construction cost, low terminal power consumption, and strong real-time characteristics, it is very suitable for γ flaw detection in the wild working environment, no power supply, weak network signal and other scenarios application.